\documentclass[12pt]{article}

\usepackage{epsfig}
\usepackage{amsmath}
\usepackage{amssymb}
\usepackage{axodraw}

\providecommand{\Dta}{{\overline{D3}}}

\begin{document}

\begin{titlepage}
\begin{flushright}
\today
\end{flushright}

\vspace{1cm}
\begin{center}
\baselineskip25pt

{\Large\bf Bekenstein-Spectrum, Hawking-Temperature and Specific
Heat of Schwarzschild Black Holes from Microscopic Chains}

\end{center}
\vspace{1cm}
\begin{center}
\baselineskip12pt {Axel Krause\footnote{E-mail: {\tt
krause@schwinger.harvard.edu}, now at Jefferson Physical
Laboratory, Harvard University, Cambridge, MA 02138, USA}}
\vspace{1cm}

{\it Physics Department,}\\[1.8mm]
{\it National Technical University of Athens,}\\[1.8mm]
{\it 15773 Athens, Greece}

\vspace{0.3cm}
\end{center}
\vspace*{\fill}

\begin{abstract}
We study the thermodynamic consequences of a recently proposed
description for a Schwarzschild black hole based on Euclidean
$(D3,D3)+(\Dta,\Dta)$ brane pairs described in terms of chain-like
excitations. A discrete mass-spectrum of Bekenstein-type is
inferred and upon identification of the black hole mass with the
chain's energy the leading corrections to both Hawking-temperature
and specific heat of the black hole are obtained. The results
indicate that for small black holes the evaporation process will
be considerably altered.
\end{abstract}

\vspace*{\fill}

\end{titlepage}

\section{Introduction}

It has been argued \cite{Bek} quite early that Quantum Gravity
should give an equidistant discrete spectrum for the horizon area
of a black hole. The logic being that the horizon area represents
an adiabatic invariant which leads to a discrete spectrum upon
quantisation with Bohr-Sommerfeld rules. Using the fact that for a
neutral, non-rotating Schwarzschild black hole the horizon area
$A_H$ is related to its mass $M_{BH}$ via its Schwarzschild radius
$r_S = 2G_4M_{BH}$, one obtains consequently the discrete
Bekenstein mass-spectrum
\begin{equation}
M_{BH} \propto \sqrt{N} \; , \quad N\in \mathbf{N}
\end{equation}
for the black hole. Such a spectrum has now been discussed and
derived in many different ways \cite{BMS}, \cite{BM}, \cite{DRY}.
In some approaches to a quantum treatment of black holes like
e.g.~the reduced phase space quantisation method \cite{DRY},
\cite{BK} this mass-spectrum gets augmented by an additional
zero-point energy. This becomes important if one addresses the
ultimate fate of the evaporating black hole but otherwise can
safely be ignored for macroscopic black holes for which $N$ will
be extremely large.

The discreteness of the mass-spectrum implies a drastic departure
from the thermal Hawking radiation spectrum \cite{BM}. Indeed
energy can only be radiated off a macroscopic black hole at
frequencies which are integer multiples of $\omega\simeq M_{Pl}^2
/M_{BH}$ and are thus in principle observable at energies much
lower than the Planck-scale. By detecting such quanta at various
energies one might be able to distinguish experimentally between
different approaches on how to quantise gravity. For instance Loop
Quantum Gravity \cite{ARS} predicts also a discrete area-spectrum
which is however not equispaced \cite{QGA}.

Here, we want to focus on a recent approach which has its roots in
String-Theory, and in which the microscopic black hole states get
identified with long chains living on the worldvolume of two dual
Euclidean brane pairs \cite{K1}, \cite{KFurther}. We will show
that this approach leads directly to a discrete Bekenstein
mass-spectrum for the D=4 Schwarzschild black hole as well.
Moreover, upon identification of the chain's energy with the black
hole's mass we will be able to derive within this approach the
leading corrections to the black hole's Hawking temperature and
its specific heat while the leading terms will coincide with the
standard results for Hawking-temperature and specific heat known
from black hole thermodynamics. For this coincidence it is
important that in the chain approach the chain's entropy is
determined unambiguously in terms of the Bekenstein-Hawking
entropy, i.e.~there is no ambiguity resulting from an undetermined
proportionality constant in this relation.

Following the proposal of \cite{K1} for the counting of
microstates of D=4 spacetimes possessing event horizons with
spherical boundary $S^2_H$ (more precisely the boundary of the
black hole, $S^2_H$, is defined as the intersection of the future
event horizon ${\cal H}^+$ with a partial Cauchy surface ending at
spatial infinity ${\cal I}^0$) one has to introduce a doublet of
Euclidean dual brane pairs $(E_1,M_1)+(E_2,M_2)$. In each pair
$E_i$ and $M_i$ are orthogonal to each other and wrap together
$S^2_H$ plus the whole internal compact space (6-dimensional for
type II String-Theory, resp.~7-dimensional for M-Theory). In the
low-energy limit where supergravity is valid $(E_1,M_1)+(E_2,M_2)$
acting as sources lead to a unique D=10/11 background solution of
D=10/11 supergravity. The D=4 spacetime mentioned above is then
identified with the D=4 external part of the D=10/11 background
solution. Thus starting from type II String-Theory it was argued
in \cite{K2} that for uncharged Schwarzschild black holes one
needs a doublet $(E_1,M_1)+({\overline E_1},{\overline M_1})$
consisting of a dual brane pair and its anti-brane equivalent.
This configuration has no charges. Moreover, to get a
non-dilatonic black hole one should take $E_1=M_1=D3$ which is the
only non-dilatonic $Dp$-brane. Further evidence for the
identification of the doublet $(D3,D3)+(\Dta,\Dta)$ with a D=4
Schwarzschild black hole was given in \cite{K2}.

\section{Chains From Branes}

Let us now explain as a specific example of the more general
proposal made in \cite{K1} the connection between the
aforementioned Euclidean branes, chain states and the entropy of
the D=4 Schwarzschild black hole. We will start with type IIB
String-Theory on a ten-dimensional Lorentzian manifold ${\cal
M}^{(1,3)}\times T^6$, i.e.~a torus compactification from ten to
four dimensions. It will be convenient to think of the $T^6$ as
the product $T^2\times T^4$. Moreover we will choose for ${\cal
M}^{(1,3)}$ the standard Schwarzschild metric solution and denote
the boundary of the D=4 black hole by $S^2_H$. We will next wrap a
Euclidean $D3$ around $S^2_H\times T^2$ and another one around the
remaining internal $T^4$. For technical reasons (in order to avoid
a mismatch overall factor of two in the derivation of the entropy)
and for physical reasons (neutral, i.e.~uncharged black holes can
only be obtained from brane-antibrane pairs whose Ramond-Ramond
(RR) charge cancels) we have to wrap in exactly the same way
another $\Dta$ around $S^2_H\times T^2$ and a second $\Dta$ around
$T^4$. Evidence that indeed the backreaction of this Euclidean
brane pair doublet can generate a D=10 background including in its
external part the D=4 Schwarzschild metric was given in \cite{K2}.
The background (as well as the brane configuration) breaks all
supersymmetry and can alternatively be characterized as a black
D6-brane in its ultra non-extreme limit. In this limit the black
D6-brane looses its magnetic RR 2-form charge while the dilaton
becomes constant thus giving a non-dilatonic vacuum solution.

Following \cite{K1}, it is then easy to see that for this set-up
of Euclidean branes the Bekenstein-Hawking (BH) entropy of the D=4
Schwarzschild black hole can be expressed purely in terms of the
Nambu-Goto actions $S_{D3},S_{\Dta}$ of the two brane pairs as
\begin{equation}
{\cal S}_{BH} \equiv \frac{A_H}{4G_4}
= (S_{D3})_{S^2\times T^2} (S_{D3})_{T^4}
+ (S_\Dta)_{S^2\times T^2} (S_{\Dta})_{T^4} \; .
\end{equation}

The crucial point now is to think of the tension $\tau_{D3}$ of a
Euclidean $D3$-brane as the inverse of a fundamental smallest
volume unit $v_{D3}$
\begin{equation}
\tau_{D3}=\frac{1}{v_{D3}}
\end{equation}
which is an interpretation more adapt to a Euclidean brane as it
treats all worldvolume dimensions equally (as opposed to an
interpretation as a mass per unit volume which allocates a special
role to the time-direction). This interpretation of a brane's
tension follows also from the `brane worldvolume uncertainty
relation' \cite{CHK} as explained in \cite{K13}, \cite{K14}.
Consequently a Euclidean $D3$-brane or likewise the
$\Dta$-antibrane can be thought of as a lattice made out of cells
with volume $v_{D3}$. The number $N_{D3}$ of such cells is then
precisely measured by the brane's Nambu-Goto action
\begin{equation}
N_{D3} = \tau_{D3}\int d^4x\sqrt{\det g} = \frac{\text{Volume of
Euclidean\,} D3}{v_{D3}}
\end{equation}
and similarly for $N_{\Dta}$. Therefore the D=4 Schwarzschild
black hole's BH-entropy becomes simply an integer $N\in\mathbf{N}$
\begin{equation}
{\cal S}_{BH} = (N_{D3})_{S^2\times T^2} (N_{D3})_{T^4}
+ (N_\Dta)_{S^2\times T^2} (N_{\Dta})_{T^4}
= N
\label{Integer}
\end{equation}
which stands for the total number of cells contained in the
combined worldvolume of the $(D3,D3)+(\Dta,\Dta)$ doublet.

In order to derive the black hole's BH-entropy by counting an
appropriate set of microstates, it was then proposed in \cite{K1}
to consider long chains\footnote{Short chains on the other hand,
composed out of two links, were used in \cite{CC1} to construct
standard model fields in warped backgrounds \cite{CC2}.} composed
out of $(N-1)$ links on the $N$ cell worldvolume lattice formed by
the $(D3,D3)+(\Dta,\Dta)$ doublet. A quantum-mechanical (a
Gibbs-correction factor was included to account for the quantum
mechanical indistinguishability of the bosonic cells) counting
then delivered an entropy for the chains
\begin{equation}
{\cal S}_c=N-\frac{1}{2}\ln N-\ln\sqrt{2\pi} + {\cal
O}\Big(\frac{1}{N}\Big) \; .
\label{ChainEntropy}
\end{equation}
By virtue of the identification (\ref{Integer}) the chains living
on the black hole's horizon thus exhibit an entropy
\begin{equation}
{\cal S}_c = {\cal S}_{BH} - \frac{1}{2}\ln {\cal S}_{BH} -
\ln\sqrt{2\pi} + {\cal O}\Big(\frac{1}{{\cal S}_{BH}}\Big)
\label{ChainEntropy2}
\end{equation}
and are therefore good candidates to explain both the black hole's
BH-entropy and the known logarithmic corrections \cite{LC1},
\cite{LC2} thereof. Note that the factor multiplying the logarithm
is $1/2$ in accordance with the results of \cite{LC2}.

\section{Bekenstein Mass Spectrum and Temperature}

Let us now see what black hole mass spectrum follows from this
proposal. By expressing ${\cal S}_{BH}$ for a Schwarz\-schild
black hole in terms of its mass $M_{BH}$
\begin{equation}
{\cal S}_{BH}=4\pi G_4M_{BH}^2
\end{equation}
one infers from the discreteness of the entropy (\ref{Integer})
that the black hole's mass-spectrum becomes discrete (quantized)
as well
\begin{equation}
M_{BH}(N) = \frac{\sqrt{N}}{\sqrt{4\pi G_4}}
\label{BekMassSpectrum}
\end{equation}
and turns out to be precisely of Bekenstein-type. This coincidence
is interesting because to arrive at this result we have only used
the geometrical interpretation of the brane's tension as the
inverse of a smallest volume unit while its standard derivation
uses the argument that the black hole's horizon area behaves as an
adiabatic invariant and can therefore be quantized according to
the Bohr-Sommerfeld rule \cite{Bek}. Notice also that our
derivation did not require the notion of the chains yet.

Before proceeding let us note that there is a very interesting
observation related to the Bekenstein spectrum
(\ref{BekMassSpectrum}). As pointed out first in \cite{BM} this
type of mass-spectrum offers an experimental verification well
below the Planck-scale. For microscopically small Planck-sized
black holes with $N$ not much bigger than one, the energy radiated
off the hole when jumping down from one energy level to the next
is of order the Planck-scale
\begin{equation}
\Delta E_N \equiv M_{BH}(N)-M_{BH}(N-1)\simeq \frac{1}{\sqrt{4\pi
G_4}}
\simeq M_{Pl} \; .
\end{equation}
However, when one considers macroscopically large black holes for
which $N\gg 1$ then a level jump is accompanied by an energy-loss
($M_\odot$ denotes the mass of the sun)
\begin{equation}
\Delta E_N \simeq\frac{1}{4\sqrt{\pi G_4 N}}
=\frac{1}{8\pi}\frac{M_{Pl}^2}{M_{BH}} =1.3\times
10^{-10}\frac{M_\odot}{M_{BH}}\,\text{eV}
\end{equation}
which can be considerably smaller than Planck-scale and therefore
possibly detectable. For instance primordial black holes with a
lifetime of order the present age of the universe have a mass
$M_{BH}=2.5\times 10^{-19}M_\odot$ and would emit quanta at an
energy of $\Delta E_N=0.5$ GeV.

Let us now come to the chains and examine their temperature in a
microcanonical ensemble approach. To this end we have to determine
the chain's energy which would ideally follow from a microscopic
Hamiltonian governing its dynamics. As this is still largely
unknown we will proceed differently. Since we know that at a
microscopic level in the approach proposed in \cite{K1} the black
hole resolves into a chain, what an observer at spacelike infinity
measures as the black hole's mass $M_{BH}$ is nothing else but the
chain's energy $E_c$. We are therefore led to identify $E_c$ with
$M_{BH}$ at leading order in $1/N$. Moreover we expect that there
could be subleading corrections in this identification at relative
level $1/N$. For instance we know that a fundamental string at
very high excitation level $n\gg 1$ can be thought of as a random
walk \cite{MT} and becomes therefore very similar to a chain.
Therefore as for the string whose energy $E\propto \sqrt{n+c} =
\sqrt{n}(1+c/2n+{\cal O}(1/n^2))$ ($c$ being a constant of ${\cal
O}(1)$ depending on the type of string one is considering)
receives subleading corrections at order $1/n$ we would expect
that also the chain energy might receive similar
corrections\footnote{Corrections of relative order $1/n$,
$n\in\mathbf{N}$ are also known to arise in other approaches which
treat the black hole's area like a harmonic oscillator and
consequently obtain a `zero-point correction' $A_H\propto(n+1/2)$
for the horizon area which translates into a $1/4n$ correction for
$M_{BH}$.}. We will therefore write ($a$ being a constant)
\begin{equation}
E_c(N) = M_{BH}(N) \Big(1+\frac{a}{N} + {\cal
O}\Big(\frac{1}{N^2}\Big)\Big)
= \frac{\sqrt{N}}{\sqrt{4\pi G_4}} \Big(1+\frac{a}{N} + {\cal
O}\Big(\frac{1}{N^2}\Big)\Big) \; .
\label{ChainEnergy}
\end{equation}

Knowing the energy and entropy for the chain then allows us to
determine the chain's temperature $T_c$ in a microcanonical
ensemble approach
\begin{equation}
\frac{1}{T_c}=\frac{\partial{\cal S}_c}{\partial E_c}
=\frac{dN}{dE_c}\frac{d{\cal S}_c}{dN} = 4\sqrt{\pi G_4}\sqrt{N} \left(
1 + \big(a-\frac{1}{2}\big)\frac{1}{N}+{\cal
O}\Big(\frac{1}{N^2}\Big)
\right)
\label{ChainTemp1}
\end{equation}
where we regard $N$ as a quasi-continuous parameter. By using the
Bekenstein mass-spectrum for the black hole, the chain temperature
can be expressed through the hole's mass as
\begin{equation}
\frac{1}{T_c}=\frac{1}{T_H}+\frac{(2a-1)}{M_{BH}}
+{\cal O}\Big(\frac{M_{Pl}^2}{M_{BH}^3}\Big)
\label{ChainTemp2}
\end{equation}
where
\begin{equation}
\frac{1}{T_H}=8\pi G_4M_{BH} = 8\pi\frac{M_{BH}}{M_{Pl}^2}
\end{equation}
is the Hawking-temperature of the Schwarzschild black hole. Thus
the chain's temperature will equal the Hawking temperature for
large $N$ (which was actually clear from the fact that at leading
order the chain's energy and entropy coincide with the standard
black hole entities) but will in general deviate from it the more
the smaller $N$ becomes and therefore the smaller the black hole's
mass $M_{BH}$ becomes. This clearly indicates that the black
hole's evaporation process will be considerably altered as
compared to the standard view once the black hole becomes
sufficiently small. Indeed the first and the second term on the
rhs of (\ref{ChainTemp2}) show opposing dependences on $M_{BH}$
such that if $a<1/2$ the chain's temperature will diverge already
at some finite $M_{BH}$ value as opposed to $M_{BH}=0$ predicted
by the Hawking-temperature formula alone. As long as this feature
is not altered but sustained by even higher order corrections
(notice that the second term in (\ref{ChainTemp2}) is of order
$M_{Pl}^2/M_{BH}^2$ as compared to the $1/T_H$ term and therefore
shows that it becomes important close to the Planck regime where
all the higher order contributions suppressed so far become
important as well), it indicates that the chain's free energy
$F_c=E_c-T_c {\cal S}_c$ actually diverges at this finite $M_{BH}$
value as well, thus signalling a phase transition. This nourishes
hope that puzzles like the black hole information puzzle might be
completely avoided in this framework if one takes corrections to
the standard black hole results into account.

\section{The Specific Heat}

Let us now similarly explore the black hole's specific heat. It is
a characteristic feature of the Schwarzschild black hole to
possess a negative specific heat. From the laws of black hole
thermodynamics this is known to be
\begin{equation}
C_{BH}=-8\pi G_4M_{BH}^2 \; .
\label{SpecHeatBH}
\end{equation}
Once more we expect to reproduce this result at leading order as
here the chain and the black hole energy and temperature coincide.
However again there will be non-trivial corrections to the leading
order standard black hole result.

For a microcanonical ensemble of chains one obtains with
(\ref{ChainEnergy}) and (\ref{ChainTemp1}) the specific heat
\begin{equation}
C_c=\frac{\partial E_c}{\partial T_c}
=\frac{dN}{dT_c}\frac{dE_c}{dN} =-2N+(3-4a)+{\cal
O}\Big(\frac{1}{N}\Big) \; .
\end{equation}
Using (\ref{BekMassSpectrum}) this can be expressed in terms of
the black hole's mass as
\begin{equation}
C_c = -8\pi \frac{M_{BH}^2}{M_{Pl}^2} + (3-4a) +{\cal
O}\Big(\frac{M_{Pl}^2}{M_{BH}^2}\Big) \; .
\end{equation}
Again the chain's specific heat coincides with the standard black
hole result at leading order but deviates from it at rather small
black hole masses as the first correction term $3-4a$ is of order
$M_{Pl}^2/M_{BH}^2$ relative to the leading order term. Depending
on whether $a>3/4$ or $a<3/4$ the specific heat would either be
negative for all masses thus pointing towards some unstoppable
instability or would become zero already at some finite mass value
$M_{BH}$ close to the Planck scale (under the premise that even
higher order corrections do not spoil this result). Therefore in
contrast to the leading order result (\ref{SpecHeatBH}) which
implies an instability down to the last stages of the black hole
evaporation process, the inclusion of the correction term
indicates in the case of $a<3/4$ that at some small but finite
mass the black hole's evaporation might cease and a stable state
be reached as the specific heat becomes positive here. It would
therefore be clearly interesting to understand the dynamics of the
chain in detail in order to get a rigorous understanding of the
final stages of the black hole evaporation.

\section*{Acknowledgments}

This work was supported through the European Community's Human
Potential Program under contract HPRN-CT-2000-00148 and through
NSF grant PHY-0099544.

\newcommand{\zpc}[3]{{\sl Z.Phys.} {\bf C\,#1} (#2) #3}
\newcommand{\npb}[3]{{\sl Nucl.Phys.} {\bf B\,#1} (#2) #3}
\newcommand{\npps}[3]{{\sl Nucl.Phys.Proc.Suppl.} {\bf #1} (#2) #3}
\newcommand{\plb}[3]{{\sl Phys.Lett.} {\bf B\,#1} (#2) #3}
\newcommand{\prd}[3]{{\sl Phys.Rev.} {\bf D\,#1} (#2) #3}
\newcommand{\prb}[3]{{\sl Phys.Rev.} {\bf B\,#1} (#2) #3}
\newcommand{\pr}[3]{{\sl Phys.Rev.} {\bf #1} (#2) #3}
\newcommand{\prl}[3]{{\sl Phys.Rev.Lett.} {\bf #1} (#2) #3}
\newcommand{\prsla}[3]{{\sl Proc.Roy.Soc.Lond.} {\bf A\,#1} (#2) #3}
\newcommand{\jhep}[3]{{\sl JHEP} {\bf #1} (#2) #3}
\newcommand{\cqg}[3]{{\sl Class.Quant.Grav.} {\bf #1} (#2) #3}
\newcommand{\grg}[3]{{\sl Gen.Rel.Grav.} {\bf #1} (#2) #3}
\newcommand{\prep}[3]{{\sl Phys.Rep.} {\bf #1} (#2) #3}
\newcommand{\fp}[3]{{\sl Fortschr.Phys.} {\bf #1} (#2) #3}
\newcommand{\nc}[3]{{\sl Nuovo Cimento} {\bf #1} (#2) #3}
\newcommand{\nca}[3]{{\sl Nuovo Cimento} {\bf A\,#1} (#2) #3}
\newcommand{\lnc}[3]{{\sl Lett.Nuovo Cimento} {\bf #1} (#2) #3}
\newcommand{\ijmpa}[3]{{\sl Int.J.Mod.Phys.} {\bf A\,#1} (#2) #3}
\newcommand{\ijmpd}[3]{{\sl Int.J.Mod.Phys.} {\bf D\,#1} (#2) #3}
\newcommand{\rmp}[3]{{\sl Rev. Mod. Phys.} {\bf #1} (#2) #3}
\newcommand{\ptp}[3]{{\sl Prog.Theor.Phys.} {\bf #1} (#2) #3}
\newcommand{\sjnp}[3]{{\sl Sov.J.Nucl.Phys.} {\bf #1} (#2) #3}
\newcommand{\sjpn}[3]{{\sl Sov.J.Particles\& Nuclei} {\bf #1} (#2) #3}
\newcommand{\splir}[3]{{\sl Sov.Phys.Leb.Inst.Rep.} {\bf #1} (#2) #3}
\newcommand{\tmf}[3]{{\sl Teor.Mat.Fiz.} {\bf #1} (#2) #3}
\newcommand{\jcp}[3]{{\sl J.Comp.Phys.} {\bf #1} (#2) #3}
\newcommand{\cpc}[3]{{\sl Comp.Phys.Commun.} {\bf #1} (#2) #3}
\newcommand{\mpla}[3]{{\sl Mod.Phys.Lett.} {\bf A\,#1} (#2) #3}
\newcommand{\cmp}[3]{{\sl Comm.Math.Phys.} {\bf #1} (#2) #3}
\newcommand{\jmp}[3]{{\sl J.Math.Phys.} {\bf #1} (#2) #3}
\newcommand{\pa}[3]{{\sl Physica} {\bf A\,#1} (#2) #3}
\newcommand{\nim}[3]{{\sl Nucl.Instr.Meth.} {\bf #1} (#2) #3}
\newcommand{\el}[3]{{\sl Europhysics Letters} {\bf #1} (#2) #3}
\newcommand{\aop}[3]{{\sl Ann.~of Phys.} {\bf #1} (#2) #3}
\newcommand{\arnps}[3]{{\sl Ann.Rev.Nucl.Part.Sci.} {\bf #1} (#2) #3}
\newcommand{\jetp}[3]{{\sl JETP} {\bf #1} (#2) #3}
\newcommand{\jetpl}[3]{{\sl JETP Lett.} {\bf #1} (#2) #3}
\newcommand{\acpp}[3]{{\sl Acta Physica Polonica} {\bf #1} (#2) #3}
\newcommand{\sci}[3]{{\sl Science} {\bf #1} (#2) #3}
\newcommand{\nat}[3]{{\sl Nature} {\bf #1} (#2) #3}
\newcommand{\pram}[3]{{\sl Pramana} {\bf #1} (#2) #3}
\newcommand{\hepth}[1]{{\tt hep-th/}{\tt #1}}
\newcommand{\hepph}[1]{{\tt hep-ph/}{\tt #1}}
\newcommand{\grqc}[1]{{\tt gr-qc/}{\tt #1}}
\newcommand{\astroph}[1]{{\tt astro-ph/}{\tt #1}}
\newcommand{\desy}[1]{{\sl DESY-Report~}{#1}}

\end{document}